\documentclass[10pt,aps,prx,showpacs,preprint,longbibliography]{revtex4-1}
\usepackage{graphicx}
\usepackage{amsmath}
\usepackage{amssymb}
\usepackage{bm}
\usepackage{color}

\newcommand{\EQ}{Eq.~}

\newcommand{\FIG}{Fig.~}

\begin{document}

\title{Emergence of event cascades in inhomogeneous networks}
\author{Tomokatsu Onaga}
\email{onaga@scphys.kyoto-u.ac.jp}
\affiliation{Department of Physics, Kyoto University, Kyoto 606-8502, Japan}
\author{Shigeru Shinomoto}
\email{shinomoto@scphys.kyoto-u.ac.jp}
\affiliation{Department of Physics, Kyoto University, Kyoto 606-8502, Japan}

\begin{abstract}
There is a commonality among contagious diseases, tweets, urban crimes, nuclear reactions, and neuronal firings that past events facilitate the future occurrence of events. The spread of events has been extensively studied such that the systems exhibit catastrophic chain reactions if the interaction represented by the ratio of reproduction exceeds unity; however, their subthreshold states for the case of the weaker interaction are not fully understood. Here, we report that these systems are possessed by nonstationary cascades of event-occurrences already in the subthreshold regime. Event cascades can be harmful in some contexts, when the peak-demand causes vaccine shortages, heavy traffic on communication lines, frequent crimes, or large fluctuations in nuclear reactions, but may be beneficial in other contexts, such that spontaneous activity in neural networks may be used to generate motion or store memory. Thus it is important to comprehend the mechanism by which such cascades appear, and consider controlling a system to tame or facilitate fluctuations in the event-occurrences. The critical interaction for the emergence of cascades depends greatly on the network structure in which individuals are connected. We demonstrate that we can predict whether cascades may emerge in a network, given information about the interactions between individuals. Furthermore, we develop a method of reallocating connections among individuals so that event cascades may be either impeded or impelled in a network. 
\end{abstract}


\maketitle

\section{Introduction}
Our life is full of cause-and-effect relationships, such that past events influence the future occurrence of events. The proliferation process has been studied using both macroscopic models, such as the epidemic model~\cite{kermack27}, and microscopic models, such as the self-exciting point process proposed by Hawkes~\cite{hawkes71a,hawkes71b}. These models have been applied to analyze not only the communication of diseases~\cite{hethcote00,keeling08,vespignani12,brockmann13,pastor-satorras15} but also urban crime~\cite{Mohler11}, human activity~\cite{goffman64,crane08,kitsak10,lerman10,romero11}, economics~\cite{ait10}, genome sequences~\cite{reynaud10}, and neuronal firing~\cite{pernice11,reynaud-bouret14}. A key quantity representing the interaction in these various phenomena is the basic reproduction ratio, which is defined as the average number of additional events induced by a single event~\cite{heffernan05}. In epidemics, a disease becomes a pandemic in a homogeneous network if the reproduction ratio is greater than unity, as in a nuclear chain reaction~\cite{anderson39,scharff46,duderstadt76}, and vanishes otherwise.

Nevertheless, the event-occurrence does not cease if individuals are stimulated in external communities or exhibit spontaneous activity. In such situations, the system may still exhibit cascades of event-occurrences intermittently, even if the reproduction ratio is smaller than the epidemic threshold, as in tweets~\cite{barabasi05,crane08,kwak10} and neuronal firings {\it in vivo}~\cite{sakata09}. The nonstationary fluctuations may be terminated by reducing the reproduction ratio further~\cite{onaga14}. Event cascades can be a nuisance in some contexts, such as when the peak-demand causes vaccine shortages~\cite{hinman06,stohr10} or heavy traffic on communication lines~\cite{huberman97}, but may be beneficial in other contexts; for example, spontaneous activity in neural networks may be used to generate motion or store memory~\cite{jaeger04,sussillo09,laje13}. Thus, it is important to comprehend the mechanism by which such cascades appear. We show that such a transition between stationary and nonstationary states generally occurs in every proliferation system, obtain the condition on which cascades may emerge in a given network, and suggest a systematic method for controlling systems to oppress or promote the event-occurrence bursts.

\section{Mean rate of event-occurrence}
Although the epidemic model and the Hawkes process appear to differ from one another, they have something in common because both were constructed to describe the proliferation processes. To identify their common features, we first revise them by considering realistic constraints. 

\subsection{Revising the epidemic model}
For an epidemic model, we consider the susceptible-infected-susceptible (SIS) model describing the situation in which infected individuals may recover without immunity:
\begin{equation}
di / dt =\beta si-\gamma i,
\end{equation}
where $i$ and $s$ are the fractions of infected and susceptible individuals, respectively ($i+s=1$); $\beta$ is the rate at which susceptible individuals are infected by contacting infected individuals; and $\gamma$ is the rate at which infected individuals recover and regain susceptibility. The infected individuals asymptotically vanish if the reproduction ratio $R_0 = \beta/\gamma$ is smaller than or equal to unity; otherwise, the fraction is finite: $i_{\infty} \equiv \lim_{t\to\infty}i(t) = 1-1/R_0$. To consider extrinsic or spontaneous activation, we suggest revising the SIS model by adding an inflow to the infected population from the susceptible one:
\begin{equation}
di / dt =\beta si-\gamma i+\rho s.
\end{equation}
\begin{figure*}[t]
\centering
\includegraphics[width=11.4cm]{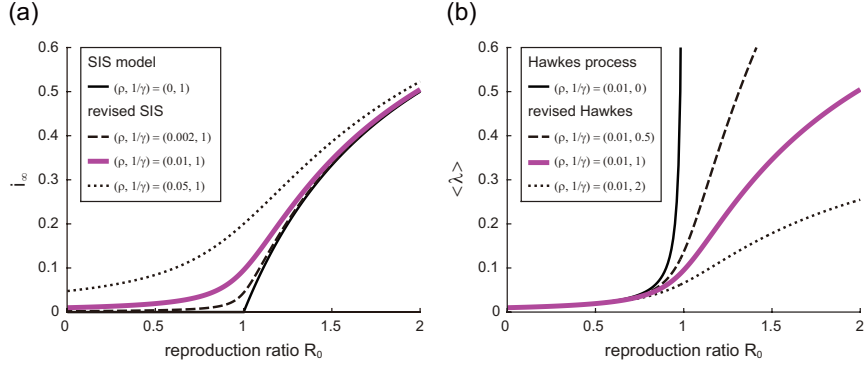}
\caption{
Mean occurrence rate obtained by the revised epidemic model and the revised Hawkes process. (a) The equilibrium fraction of infected individuals $i_{\infty}$ obtained by the susceptible-infected-susceptible (SIS) model revised by considering spontaneous activity $\rho>0$. (b) The mean occurrence rate $\langle \lambda \rangle$ of the Hawkes process revised by introducing the refractory period $1/\gamma >0$. These models give identical equilibria, $\langle \lambda \rangle = \gamma i_{\infty}$, for the same spontaneous activation $\rho$ and the refractory period $1/\gamma$ (magenta lines in (a) and (b)).
}
\label{f1}
\end{figure*}
In the presence of spontaneous activity $\rho (>0)$, the asymptotic fraction of infected individuals remains positive and smoothly increases with $R_0$, and, accordingly, the epidemic transition at $R_0=1$ is softened (\FIG\ref{f1}(a)).

\subsection{Revising the Hawkes process}
The Hawkes process considers spontaneous occurrences in terms of the positive base rate $\rho$ and describes the manner in which the event-occurrence rate $\lambda(t)$ is modulated by past events:
\begin{equation}
\lambda (t)=\rho+R_0 \sum_{k}h(t-t_k),
\label{hawkes}
\end{equation}
where $t_k$ is the occurrence time of the $k$th event. The history kernel $h(t)$ satisfies the causality, $h(t)=0$ for $t<0$, and the normalization, $\int_{0}^{\infty}h(t)dt=1$. By taking the ensemble average, the average rate of event-occurrence is obtained as $\langle \lambda(t) \rangle=\rho /(1-R_0)$. The divergence at $R_0 = 1$ arises from instantaneous reactivation, which is an artifact caused by the simplicity of the linear model, which ignores the refractory period during which each individual does not recover susceptibility. Here, we suggest revising the model by introducing the effect of a refractory period $1/\gamma$:
\begin{eqnarray}
\lambda (t)=\left(1- \frac{ \lambda (t) }{\gamma} \right) \left(\rho+R_0 \sum_{k}h(t-t_k) \right).
\label{hawkes_rev}
\end{eqnarray}
By taking an ensemble average and approximating $\langle \lambda^2 \rangle$ with $\langle \lambda \rangle^2$, we obtain the average rate of event-occurrences as $\langle \lambda \rangle = \gamma i_{\infty}$, where $i_{\infty}$ is the asymptotic fraction obtained for the revised SIS model (\FIG\ref{f1}(b)). Thus, the epidemic model and the Hawkes process may represent the identical mean occurrence rate by taking the spontaneous activation and the refractory periods into account.

\section{Fluctuations in the event-occurrence}
Although the systems no longer exhibit a clear transition at $R_0=1$ causing the catastrophic chain reaction, they may still show nonstationary fluctuations with intermittent cascades of event-occurrences at lower reproduction ratios; when the reproduction ratio is even smaller, they may remain stationary, producing apparently random events over time.

\subsection{Epidemic Markov process}
\begin{figure*}[t]
\centering
\includegraphics[width=11.4cm]{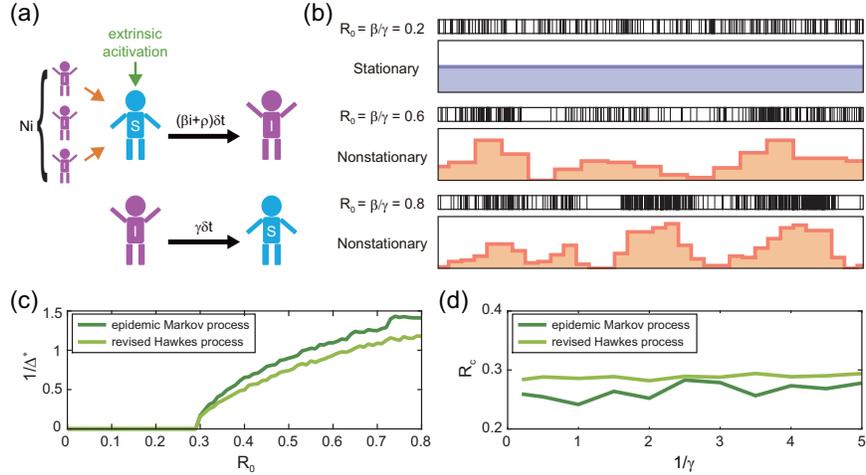}
\caption{
Stationary-nonstationary (SN) transition in the epidemic Markov process and the revised Hawkes process. (a) Microscopic epidemic Markov process. (b) Sample sequences of infected times obtained by the epidemic Markov processes (model parameters: $\beta=0.06$, $0.18$, and $0.24$, with $\gamma=0.3$, $\rho=0.5$, and $N=1000$). Below the raster diagrams are the fitted optimal histograms. (c) Inverse optimal binsize $1/\Delta^*$ plotted against the reproduction ratio $R_0$ ($\gamma=0.3$, $\rho=0.5$, and $N=1000$). (d) Critical points of the SN transition $R_c$ for the epidemic Markov and revised Hawkes processes, plotted against the refractory period $1/\gamma$.
}
\label{f2}
\end{figure*}
The SIS model that addresses the mean population dynamics cannot represent fluctuations in the event-occurrences. Here, we construct and simulate a Markov process of microscopic dynamics in which individuals become infected and recover to be susceptible (\FIG\ref{f2}(a)). In every interval of a small time-step $\delta t$, each susceptible individual in total population N may become infected with a probability $(\beta i + \rho) \delta t$, which shifts as $i \to i+1/N$. Each infected individual may regain their susceptibility with the probability $\gamma \delta t$, which shifts as $i \to i-1/N$. Figure~\ref{f2}(b) depicts sample sequences of infected times: At a reproduction ratio of $R_0 = \beta/\gamma = 0.2$, the occurrence of infection appears random across individuals but stationary in the whole system. By contrast, at $R_0=0.6$ or $0.8$, the event sequence appears nonstationary, exhibiting spontaneous cascades of occurrences. Indeed, microscopic fluctuation is amplified to be macroscopically visible, whereas the average rate is restrained to be finite.

\subsection{Defining the stationarity of a given sequence}
We suggest deciding whether a given series of events is stationary or nonstationary based on whether proper rate estimators conclude a constant rate or a fluctuating rate, respectively. Here, we adopt a method of selecting the histogram bin size to minimize the expected mean square error between the histogram and the unknown underlying rate~\cite{shimazaki07}. Note that the decision regarding stationary vs nonstationary state is common across proper rate estimators, such as the Empirical Bayes and variational Bayes Hidden Markov estimators~\cite{shintani12}.

Histograms were fitted to the data: The optimal bin size $\Delta^*$ was diverging (becoming as large as the entire observation period) when $R_0=0.2$, whereas it was finite (significantly smaller than the entire period) when $R_0=0.6$ and $0.8$. Figure~\ref{f2}(c) depicts the manner in which the inverse bin size $1/\Delta^*$ varies with the reproduction ratio $R_0$. $1/\Delta^*$ remains near zero if $R_0$ is smaller than approximately $0.3$, and it departs from zero otherwise, thus exhibiting the stationary-nonstationary (SN) transition.

\subsection{Revised Hawkes process}

We also simulated the revised Hawkes process (\ref{hawkes_rev}) with parameters identical to those used for the epidemic Markov process. Here, events are indicated by repeating the Bernoulli trials with a probability of $\lambda (t) \delta t$ in every small interval of $\delta t$. By plotting $1/\Delta^*$ versus $R_0$, we can also observe the transition (\FIG\ref{f2}(c)). Figure~\ref{f2}(d) depicts the critical reproduction ratios obtained for the epidemic Markov process and the revised Hawkes process plotted against the refractory period $1/\gamma$, indicating that the SN critical points for both models are robustly close to $1-1/\sqrt{2} \approx 0.3$, which was obtained for the original (linear) Hawkes process ($1/\gamma=0$) in our previous study~\cite{onaga14}. Note that the critical point is independent of the shape of the kernel, as well as the base rate.

\section{Event-occurrences in inhomogeneous networks}
Finally, we consider the emergence of cascades in a population of individuals interacting through inhomogeneous connections. 

\subsection{Multivariate Hawkes process}
\begin{figure*}[t]
\centering
\includegraphics[width=11.4cm]{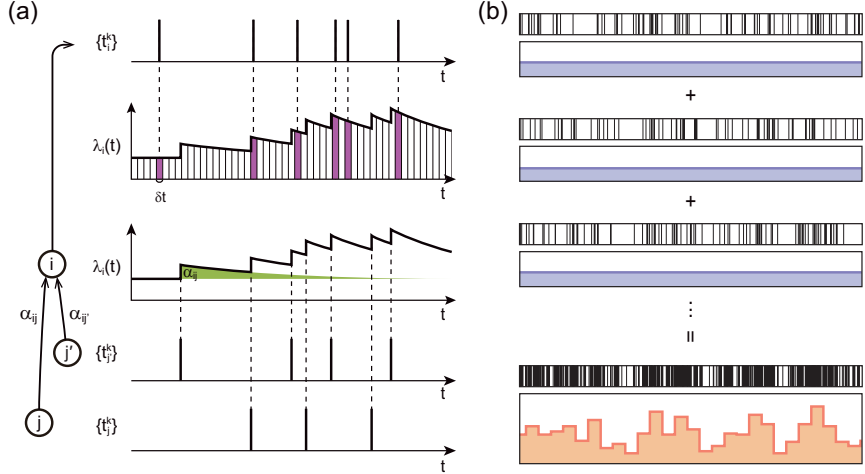}
\caption{
Multivariate Hawkes process. (a) The rate of event-occurrences in each node is modulated by the influence of events generated at other nodes, and events are derived from the underlying rate $\lambda_i(t)$. (b) The manner in which the nonstationary fluctuations become visible by superposing event series in individual nodes.
}
\label{f3}
\end{figure*}

To obtain the condition for the SN transition analytically, we analyze the linear multivariate Hawkes processes (\FIG\ref{f3}(a)). The rate of event-occurrences in the $i$th individual or node ($i=1, 2, \cdots, N$) is given as
\begin{equation}
\lambda_i(t)=\rho_i+\sum_{j=1}^N \alpha_{ij} \sum_k h (t-t_j^k),
\label{mvhawkes}
\end{equation}
where $\rho_i$ represents the base rate, $t_j^k$ is the occurrence time of the $k$th event in the $j$th node, and $\alpha_{ij}$ represents the interaction from the $j$th node to the $i$th node. Because of the interactions between individuals, $\bm{A} \equiv \{ \alpha_{ij} \}$, the average rates $\langle \bm{\lambda} \rangle = \{\langle \lambda_i \rangle\}$ are shifted from the base rates $\bm{\rho} = \{\rho_i\}$ as $\langle \bm{\lambda} \rangle = \bm{L} \bm{\rho}$, where $\bm{L}$ is the Leontief inverse~\cite{waugh50}: $\bm{L} \equiv \sum_{n=0}^{\infty} \bm{A}^n =\bm{I}/\left( \bm{I}-\bm{A}\right)$.

\subsection{Controlling the event-occurrence cascades}
Even when the fluctuations are not detectable at any single node, all events occurring in the entire set of nodes may exhibit visible fluctuations because the signal-to-noise ratio may increase when multiple series of events are superposed (\FIG\ref{f3}(b)). By analyzing equation \EQ(\ref{mvhawkes}), we obtain the condition required for a series of events occurring in an entire population to be nonstationary (see Appendix \ref{appendix:a}):
\begin{equation}
C \equiv \frac{\sum_{i,j}\left( \bm{L} \bm{\Lambda} \bm{L}^T \right)_{ij} }{ \sum_{i} \langle \lambda_i \rangle } > 2,
\label{conditionall}
\end{equation}
where $\bm{\Lambda} \equiv {\rm diag}\left(\langle \bm{\lambda} \rangle \right)$. 

Here, we consider 0-1 connectivity with strength $\alpha_{ij} = 0$ or $R_0/Nc$, where $c$ is the fraction of connections. For a fully connected network $c=1$, the summed rate $\lambda (t) = \sum_{i=1}^{N} \lambda_i (t)$ obeys the original Hawkes process \EQ(\ref{hawkes}) with $\rho = \sum_{i=1}^{N} \rho_i$. For the Erd\H{o}s-R\'{e}nyi network, in which a pair of nodes is independently randomly linked at a fraction of $c$, the SN critical point remains near $0.3$ for a wide range of $c$.

\begin{figure*}[t]
\centering
\includegraphics[width=17cm]{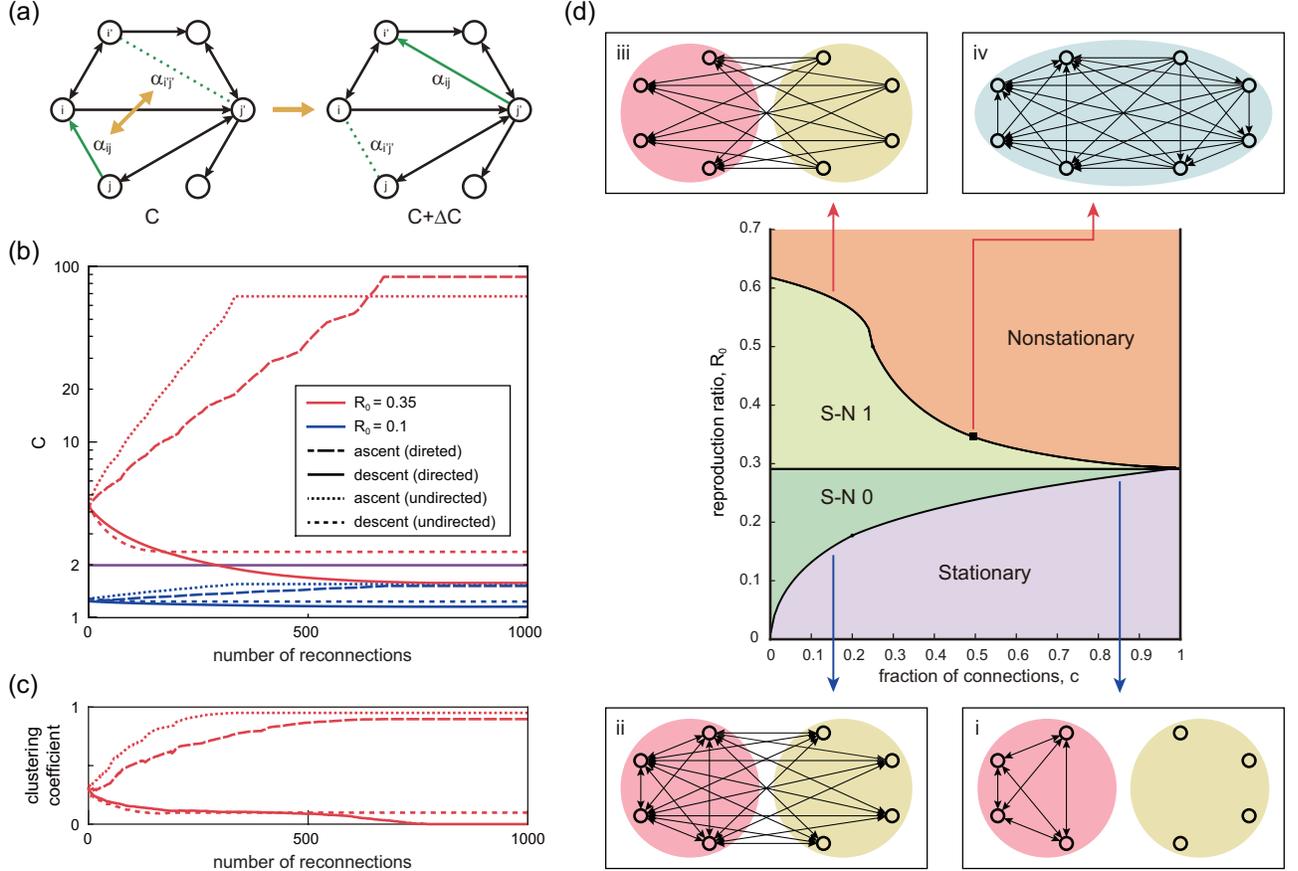}
\caption{
Controlling the emergence of event-occurrence cascades. (a) Elementary process of exchanging connections $\alpha_{ij}$ and $\alpha_{i'j'}$. (b) The manner in which the potential for the cascades $C$ is altered by the steepest ascent or descent based on \EQ(\ref{DC}), starting from the Erd\H{o}s-R\'{e}nyi networks. The blue and red lines represent the cases of $(R_0, c) = (0.1, 0.1)$, and ($0.35, 0.1$), respectively ($N = 100$). (c) Changes in the average clustering coefficient according to the reallocation of connections when $(R_0, c) = (0.35, 0.1)$. (d) Parameter ranges of ($R_0, c$) in which the networks may be either stationary or nonstationary. Some solvable extreme configurations that give low and high critical points in $R_0$ are depicted.
}
\label{f4}
\end{figure*}

Nevertheless, it is possible to shift $C$ by reallocating the connections between individuals; exchanging connections $\alpha_{ij}$ and $\alpha_{i'j'}$ may alter $C$ by
\begin{equation}
\Delta C = -\frac{\alpha_{ij} H_{ij}-\alpha_{i'j'} H_{i'j'}}{\sum_k\langle\lambda_k\rangle} + \mathcal{O}(R_0/N)^2,
\label{DC}
\end{equation}
where $H_{ij} \equiv \left(\left( \sum_k L_{ki} \right)^2-C\right)\rho_j + 2\left( \sum_k L_{kj} \right)\left\langle\lambda_j\right\rangle$. To raise or lower $C$, we repeat exchanging a pair of connections that maximizes or minimizes $\Delta C$ (\FIG\ref{f4}(a)).

A network of $(R_0, c)$ may change the state from stationary to nonstationary if $C$ steps across the critical value of $2$ from below or vice versa by reconnecting individuals. Figure~\ref{f4}(b) demonstrates the manner in which $C$ is altered by the steepest ascent or descent based on \EQ(\ref{DC}). When $(R_0, c)=(0.1, 0.1)$, $C$ remains below 2, even when all connections are reallocated; however, when $(R_0, c)=(0.35, 0.1)$, $C$ exceeds 2, indicating that the system may change between nonstationary ($C>2$) and stationary ($C\le2$). 

\subsection{Network structures favorable for inciting or impeding cascades}
In the study of epidemics, whether the epidemic threshold is higher in the clustered networks~\cite{keeling99,miller09,pastor-satorras15} or not ~\cite{newman03} has been controversial. Here, we are not addressing the epidemic transition, but we are interested in how the clustering of individuals influences the S-N transition. Figure~\ref{f4}(c) depicts the manner in which the average clustering coefficient changes with our gradient ascent or descent of $C$, indicating that clustering tends to facilitate the event-occurrence cascades. A similar tendency was reported in neural network simulations~\cite{laje13}. An advantage of our method is that we can control the cascade bursting activity by systematically rearranging connections based on a single measure: $C$.

Figure~\ref{f4}(d) depicts the range of $(R_0, c)$ in which a network may exhibit stationary and nonstationary states. In the ``Stationary'' regime, the systems never generate visible cascades, even when connections are reallocated, whereas in the ``Nonstationary'' regime, the systems always exhibit cascades. In the ``S-N 0'' regime, networks may be either stationary or nonstationary, depending on the manner in which individuals are connected. In the ``S-N 1'' regime, networks may be either stationary or nonstationary if the reciprocal connections can be controlled independently, whereas undirected networks, whose connections are reciprocally symmetric $\{\alpha_{ij}=\alpha_{ji}\}$, remain nonstationary, always generating cascades of events. For undirected networks, the critical point $R_0$ appears to be bounded at approximately $0.3$.

As extreme configurations with low critical points in $R_0$, we considered networks of (i): the nodes of one group are fully connected within the group, and the others are isolated; and (ii): the nodes of one group are fully connected within the group and also receive undirected (reciprocal) connections from all nodes of another group. The critical points of $R_0$ exhibit crossover at $c=0.20$, and above and below this point, those of (i) and (ii) are lower. The criticality in low $R_0$ implies that these configurations tend to facilitate cascade bursting. As an opposite extreme, we considered another configuration (iii): each node of one group receives only directed connections from another group. We also considered a specific hierarchical networks of (iv): every node exerts the influence over the lower hierarchy nodes in a manner that connections form a triangular matrix. All critical points for these specific cases are obtained analytically (see Appendix \ref{appendix:b}) and plotted in \FIG\ref{f4}(d): The configurations (i) and (ii) tend to incite bursting, whereas the configurations (iii) and (iv) impede bursting. These observations imply that networks in which small number of individuals occupy reciprocal connections favor event cascade bursting, whereas directed unilateral connections tend to impede cascades. 

\begin{figure*}[t]
\centering
\includegraphics[width=11.4cm]{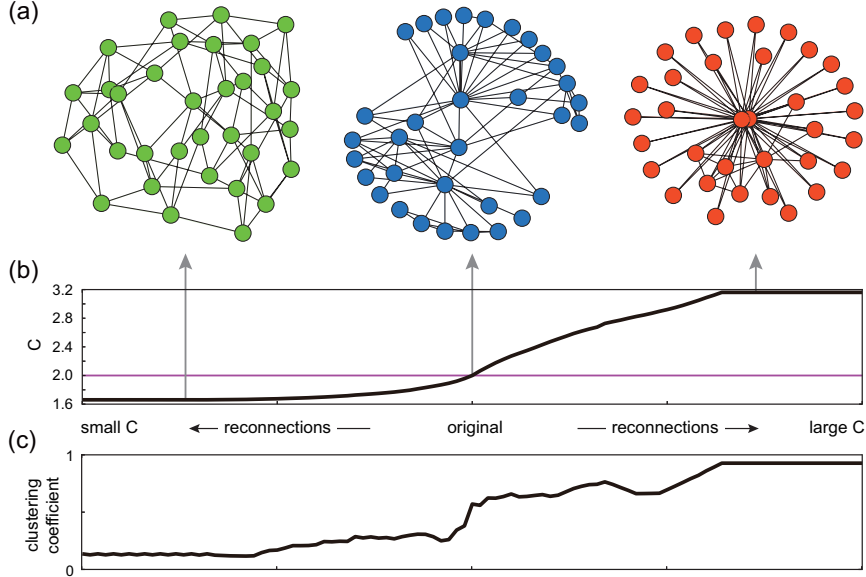}
\caption{
Reallocating connections between individuals. (a) The manner in which the (undirected) friendships between 34 people in “Zachary's Karate Club” are reconnected by increasing or decreasing $C$, respectively for inciting or impeding cascades of communications. (b) The change in the potential for the cascades $C$ with the steepest ascent (rightward) or descent (leftward) based on \EQ(\ref{DC}), (c) Average clustering coefficient.
}
\label{f5}
\end{figure*}

Finally we demonstrate the manner in which the network is reconnected by our method for inciting or impeding cascades (\FIG\ref{f5}); we take up the network of ``Zachary's Karate Club''~\cite{zachary77} and reallocate the (undirected) friendships between 34 people. By assuming that the original network is at the criticality $C=2$, we reorganized the network so that their entire communications becomes stationary or nonstationary. The operation to impede the communication cascades ($\Delta C<0$) tended to make all people be connected harmoniously. Contrariwise, the operation to incite the cascade ($\Delta C>0$) made the society polarized into two kinds of people; a few people have exclusive rich connections, while others are connected only with those people. 

\section{Discussion}
Here, we showed that the proliferation process may exhibit the SN transition at which nonstationary cascades of event-occurrences emerge from the stationary process. The critical reproduction ratio for the SN transition is much smaller than the threshold for the conventional epidemic transition exhibiting chain reactions. The SN transition may depend greatly on the manner in which individuals are connected. We developed a theory for predicting the occurrence of cascades in a network. We also suggested a method of reconnecting individuals for impeding or inciting cascade bursting in a network. To set about applying our theory to real-world problems, we need information regarding interactions. It may be the next challenge to analyze subthreshold dynamics occurring in a variety of proliferation processes and develop methods of estimating interactions between nodes or individuals from their sequences of events.

\section*{Acknowledgments}
We thank Takaaki Aoki, Hiroya Nakao, and Naoki Masuda for their advice and comments. This study was supported in part by a Grant-in-Aid for Scientific Research to S.S. from MEXT Japan (26280007) and by JST and CREST. T.O. is supported by a JSPS Research Fellowship for Young Scientists.

\appendix

\section{Derivation of the condition for the SN transition}\label{appendix:a}
Here, we derive the condition for the SN transition for a given series of events. It has been proven that the optimal bin size may be finite if fluctuation in the underlying rate $\delta\lambda (t) \equiv \lambda (t) - \langle \lambda \rangle$ satisfies the condition~\cite{koyama04,shintani12}
\begin{equation}
\frac{1}{\langle \lambda \rangle }\int_{-\infty}^{\infty}\langle \delta\lambda (t+s)\delta\lambda (t)\rangle ds >1,
\label{SN}
\end{equation}
and diverges otherwise. This is derived as follows. The mean square error between the underlying rate $\lambda(t)$ and the histogram $\hat{\lambda}(t)$ is given as
\begin{equation}
S=\lim_{T \to \infty}\frac{1}{T}\int_0^T \left\langle \left(\lambda(t)-\hat \lambda(t)\right)^2 \right\rangle dt,
\label{mise}
\end{equation}
where $T$ is the entire observation interval, and the bracket represents the ensemble average over the possible realization of the stochastic process. In each bin of size $\Delta$, the histogram $\hat\lambda (t)$ is a constant whose height is the number of events $K$ divided by the bin size $\Delta$. Thus, the mean square error is transformed as
\begin{equation}
S=\left\langle \frac{1}{\Delta} \int_0^{\Delta} \left( \lambda^2(t) -\frac{2 K}{\Delta} \lambda(t) + \frac{K^2}{\Delta^2} \right) dt \right\rangle.
\label{mise1}
\end{equation}
The expected number of events in each interval is given by integrating the underlying rate: $\left\langle K \right\rangle=\int_0^{\Delta} \lambda(t) dt$. Because events are independently drawn, the Poisson relation holds: $\left\langle K^2 \right\rangle=\left\langle K \right\rangle^2+\left\langle K \right\rangle$. Inserting these relations into \EQ(\ref{mise1}), we have
\begin{equation}
S=\phi(0)+\frac{\left\langle \lambda \right\rangle}{\Delta}-\frac{1}{\Delta^2}\int_0^{\Delta}dt\int_{-t}^t \phi(s) ds,
\end{equation}
where $\phi(s) \equiv \left\langle \lambda(t+s)\lambda(t) \right\rangle-\left\langle \lambda \right\rangle^2$ is the correlation of the rate fluctuation or
$\phi(s) = \left\langle \delta \lambda(t+s) \delta \lambda(t) \right\rangle$,
where $\delta \lambda(t) \equiv \lambda(t) - \left\langle \lambda \right\rangle$ is the temporal fluctuation of the rate. The mean square error may have a minimum at some finite $\Delta$. Based on the second-order transition in which the minimum position $\Delta^*$ goes to infinity or $1/\Delta^*$ goes to zero continuously, the condition for the transition is given as
\begin{equation}
\left. \frac{dS}{d(1/\Delta)}\right|_{\Delta=\infty} < 0.
\end{equation}
This can be summed up as a condition of the rate fluctuation given in inequality (\ref{SN}) if $\int_0^{\infty}s \phi(s) ds$ is finite. This condition was derived from the optimization of a histogram and was found to be identical to that derived from the marginal likelihood maximization of the Bayesian rate estimator, implying that this condition may be a universal bound for detecting rate fluctuation~\cite{koyama07}.

For the linear self-exciting point process, the power spectrum of the rate fluctuation or the Fourier transformation of the autocorrelation $\phi(s) \equiv \langle \delta \lambda (t+s) \delta \lambda (t) \rangle$ was obtained by Hawkes~\cite{hawkes71a}. The result can be summarized as~\cite{onaga14}
\begin{equation}
\tilde\phi_{\omega} = \left( \frac{1}{(1-R_0 \tilde h_{\omega})(1-R_0 \tilde h_{-\omega})} - 1 \right)\langle \lambda \rangle,
\end{equation}
where $\tilde h_{\omega}$ is the Fourier transform of the kernel function $h(t)$. Because $\tilde\phi_{0} = \int_{-\infty}^{\infty}\phi(s) ds = \int_{-\infty}^{\infty}\langle \delta\lambda (t+s)\delta\lambda (t)\rangle ds $, the condition for the linear self-exciting process to be nonstationary is obtained as $1 / (1-R_0)^2 > 2$. Thus, the SN transition occurs at $R_0=1-1/\sqrt{2}$ independent of the time course of supplementary probability $h(t)$ and the base rate $\rho$.

The multivariate Hawkes process (\ref{mvhawkes}) is also analytically tractable; in particular, the Fourier zero-mode of the correlations $\bm{\phi}(s) \equiv \{ \phi_{ij}(s) \} \equiv \{\langle \delta \lambda_i(t+s) \delta \lambda_j (t) \rangle \}$ is obtained as~\cite{hawkes71b,onaga14},
\begin{equation}
\tilde{\bm{\phi}}_0 = \bm{L} \bm{\Lambda} \bm{L} ^T- \bm{\Lambda}.
\end{equation}
Each node may exhibit the SN criticality if the correlation of each individual, $\phi_{ii}(s)$, satisfies the SN condition. Even when rate fluctuations are not detectable at any single node, the summed activity of multiple nodes may exhibit fluctuation. The condition for the superposed series to exhibit the SN transition is obtained by applying the nonstationary condition (\ref{SN}) to the summed rate $\lambda (t) = \sum_{i=1}^{N} \lambda_i (t)$, thus leading to inequality (\ref{conditionall}).

\section{Criticality conditions for extreme configurations}\label{appendix:b}
In the following, we give the criticality conditions $C=2$ in \EQ(\ref{conditionall}) for the proposed extreme configurations (i), (ii), (iii), and (iv) (\FIG\ref{f4}(d)). The connectivity of configuration (i) is given as
\begin{eqnarray}
A=\frac{R_0}{Nc}\left(\begin{array}{cc}
\bm{1}_{M,M}&\bm 0_{M,N-M}\\
\bm 0_{N-M,M}&\bm 0_{N-M,N-M}
\end{array}\right), 
\end{eqnarray}
where $\bm{1}_{n,m}$ and $\bm{0}_{n,m}$ are $n \times m$ matrices consisting of all elements of $1$ and $0$, respectively. In this case, the fraction of connections is related to $M$ as $c=M^2/N^2$, and the critical point is obtained by solving a cubic equation:
\begin{equation}
(x-c^{1/2})^3= c^{1/2} x^3 - c x^2- c^{3/2} x.
\end{equation}

The connectivity of the configuration (ii) is given as
\begin{eqnarray}
A=\frac{R_0}{Nc}\left(\begin{array}{cc}
\bm{1}_{M,M}&\bm{1}_{M,N-M}\\
\bm{1}_{N-M,M}&\bm 0_{N-M,N-M}
\end{array}\right). 
\end{eqnarray}
The critical point is obtained by solving a fifth-degree equation:
\begin{eqnarray}
&&2b^3 x^5 + 2b^2(3-2b)(1+b)x^4 + b (6-12b+5b^2)(1+b)^2x^3 \nonumber\\ 
&&+ (1-b)(2-10b+b^2)(1+b)^3x^2 - (1-b)^2(4+b)(1+b)^4x \nonumber\\ 
&&+ (1-b)^2(1+b)^5 =0, 
\end{eqnarray}
where $b=\sqrt{1-c}=(N-M)/N$.

The connectivity of the configuration (iii) is given as
\begin{eqnarray}
A=\frac{R_0}{Nc}\left(\begin{array}{cc}
\bm 0_{N-M,N-M}&\bm 0_{N-M.M}\\ 
\bm{1}_{M,N-M}&\bm 0_{M,M}
\end{array}\right). 
\end{eqnarray}
In this case, $c$ is related to $M$ with $c=M(N-M)/N^2$, and the critical point is obtained as
\begin{equation}
R_0=\frac{1+\sqrt{1-4c}}{4}\left(-1+\sqrt{1+2c^{-1}\left(1-\sqrt{1-4c}\right)}\right).
\end{equation}

The connectivity of the configuration (iv) is given by the triangular matrix,
\begin{eqnarray}
A=\frac{R_0}{Nc}\left(\begin{array}{cccc}
0&0&\cdots&0\\
1&\ddots&\ddots&\vdots\\
\vdots&\ddots&\ddots&0\\
1&\cdots&1&0\\
\end{array}\right). 
\end{eqnarray}
In this case, $c=\lim_{N\to\infty}(N-1)/(2N)=1/2$, and the critical point is obtained as
\begin{equation}
R_0=\lim_{N\to\infty}\frac{N-1}{2}\left(2^{1/(N-1)}-1\right)=\frac{\log 2}{2}\approx0.35
\end{equation}

%

\end{document}